\documentclass[conference]{IEEEtran}
\usepackage{amsmath}
\usepackage{graphicx}

\newcommand{\isArxiv}

\ifdefined\isArxiv
\usepackage{fancyhdr}
\fancypagestyle{firstpage}{%
  \rhead{Article under review for IEEE GLOBECOM ’22}
}
\fi

\begin{document}

\title{A Finite-Range Search Formulation\\ of Maximum Likelihood MIMO Detection\\ for Coherent Ising Machines}

\author{\IEEEauthorblockN{Abhishek~Kumar~Singh$^{1}$, Davide~Venturelli$^{2}$, and Kyle~Jamieson$^1$}\\
\IEEEauthorblockA{$^1$\textit{Department of Computer Science, Princeton University}\\$^2$\textit{USRA Research Institute for Advanced Computer Science}}}

\maketitle
\ifdefined\isArxiv
\thispagestyle{firstpage}
\fi
\begin{abstract}
The last couple of years have seen an emergence of physics-inspired computing for maximum likelihood MIMO detection. These methods involve transforming the MIMO detection problem into an Ising minimization problem, which can then be solved on an Ising Machine. Recent works have shown promising projections for MIMO wireless detection using Quantum Annealing optimizers and Coherent Ising Machines. While these methods perform very well for BPSK and 4-QAM, they struggle to provide good BER for 16-QAM and higher modulations. In this paper, we explore an enhanced CIM model, and propose a novel Ising formulation, which together are shown to be the first Ising solver that provides significant gains in the BER performance of large and massive MIMO systems, like $16\times16$ and $16\times32$, and sustain its performance gain even at 256-QAM modulation. We further perform a spectral efficiency analysis and show that, for a $16\times16$ MIMO with Adaptive Modulation and Coding, our method can provide substantial throughput gains over MMSE, achieving $2\times$ throughput for SNR $\leq25$ dB, and up to $1.5\times$ throughput for SNR $\geq 30$ dB.
\end{abstract}

\IEEEpeerreviewmaketitle

\section{Introduction}
Even after two decades of research on MIMO detection, the trade-off between performance and feasibility is still unresolved~\cite{mimoSurveyLarge}. On one extreme, there are methods like the Sphere Decoder~\cite{sphereIeee} that provide optimal performance, but become computationally infeasible for large systems due to exponential complexity~\cite{sphereComp}. On the other extreme, linear detectors~(MMSE, Zero Forcing) meet the fast computational requirements of real-world systems, but suffer from severe performance degradation in scenarios where the number of users approaches the number of base station antennas. A myriad of methods lie between these two extremes and try to achieve an acceptable trade-off between performance and complexity; however, finding a near-optimal MIMO detector that can meet the processing constraints of large real-world wireless systems remains a crucial problem.  

The past few years have seen an ever-growing interest of researchers in physics-inspired computation. Several computation technologies like Quantum Annealing~(QA), Coherent Ising Machines~(CIM), Oscillator-based Ising Machines~(OIM), \textit{etc.} have been proposed in the literature and have shown promising results for solving several challenging computational problems~\cite{mohseni2022ising}. The Maximum-Likelihood MIMO detection~(ML-MIMO) is known to be NP-Hard, and the complexity of optimal methods scales exponentially with the size of the system, making them infeasible for real-life scenarios. Physics-inspired Ising-machine-based computation is thus an exciting alternative to conventional MIMO detection methods, with several works demonstrating the potential of Ising machine-based MIMO detection~\cite{ri-mimo,minsung,minsungParallelTemp,pSuccessQA}. While these works demonstrate promising results, they have also reported a decline in performance with an increase in the modulation order. 

We believe that Ising-machine-based MIMO detectors could bridge the gap between optimal performance and feasibility for practical systems. However, improving the Ising-machine-based MIMO detector's performance for higher modulations than what was done in previous work is critical for achieving this goal. In this paper, we propose a novel Ising formulation of the ML-MIMO problem. We also explore an enhanced model for the Coherent Ising Machines, based on an effect discovered in \cite{leleu2019destabilization} and further developed for CIMs in multiple other works, which enhances the probability of finding the optimal solution by modifying the CIM dynamics to avoid local minima and limit cycles. 

The existing literature relies on either a direct representation of the QAM constellation using \textit{spin} variables~\cite{minsung} (that take values $-1$ or 1) or the regularized Ising formulation~\cite{ri-mimo} to transform the ML-MIMO problem into an Ising problem. In our Ising formulation, we first compute a polynomial-time guess for the ML-MIMO problem using methods like MMSE, SIC~\cite{sicMimo1}, or Fixed Complexity Sphere Decoder~\cite{fcsd}. We then reformulate the ML-MIMO problem as a minimization problem that tries to find the optimal correction to the polynomial-time guess; instead of optimizing to find the most likely transmitted symbols. The motivation of this approach stems from the fact that, especially at mid or high SNR, the polynomial-time guess will not be far from the optimal solution. Hence, it makes sense to use the polynomial-time guess as the origin, and search around it to find the difference between the optimal solution and the guess; instead of searching across all possible solutions. Finally, we transform the new optimization problem into an Ising problem, that can be solved using the enhanced CIM model, and control the search radius by varying the transform parameters. 

We show that the execution of the enhanced CIM model on our novel formulation can significantly outperform the existing state-of-the-art and provide good BER performance, even for very high modulation orders like 256-QAM. We demonstrate that our methods are capable of mitigating the degradation in BER performance (for higher modulations) of Ising machine-based MIMO detectors. We evaluate the BER performance of our methods for $16\times16$ large MIMO and $16\times32$ massive MIMO systems at 4-, 16-, 64-, and 256-QAM modulation. We show that our novel DI-MIMO can achieve large performance gains over RI-MIMO~\cite{ri-mimo} and MMSE. We perform an empirical evaluation of spectral efficiency for $16\times16$ and $16\times32$ MIMO systems that use Adaptive Modulation and Coding~(AMC). We show that for a $16\times16$ MIMO with AMC, our method can provide substantial throughput gains over MMSE, achieving $2\times$ throughput for SNR $\leq25$ dB, and up to $1.5\times$ throughput for SNR $\geq 30$ dB. We further show that, for a $16\times32$ MIMO system, DI-MIMO vastly outperforms MMSE by achieving $2\times$ throughput in the mid-SNR regime ($\approx 7.5$ dB) and $1.5\times$ throughput in the high-SNR regime~($\approx 12$ dB).

The rest of the paper is organized as follows. Section~\ref{sec:related} provides a survey of the related literature. Section~\ref{sec:sysModel} describes the MIMO system model and Maximum-Likelihood MIMO detection~(ML-MIMO). Section~\ref{sec:cim} talks about Coherent Ising Machines and the amplitude heterogeneity correction-based enhancement. Section~\ref{sec:design} describes our novel "Delta Ising" formulation for ML-MIMO detection~(DI-MIMO). Section~\ref{sec:eval} contains evaluation of our methods and comparison with the existing state-of-the-art for CIM based MIMO detection~(RI-MIMO~\cite{ri-mimo}), and MMSE. We evaluate these methods' BER and spectral efficiency for large and massive MIMO configurations. Section~\ref{sec:conclusion} has concluding remarks from the authors.

\section{Related Works}
\label{sec:related}
The Sphere Decoder~\cite{sphereIeee} computes the optimal solution to the ML-MIMO problem and achieves the optimal BER, but it has exponential complexity~\cite{sphereComp} and is thus not feasible for practical systems with a large number of users and antennas. Many algorithms try to balance the performance-complexity tradeoff and provide good BER performance at a polynomial complexity. Linear detectors like MMSE and Zero-Forcing~(ZF) rely on channel inversion, while the Fixed-Complexity Sphere Decoder~\cite{fcsd} drastically reduces the complexity of the Sphere Decoder by aggressively pruning its search tree. Successive Interference Cancellation~(SIC)~\cite{sicMimo1} based techniques focus on sequentially decoding each user and canceling interference. There are several other approaches to MIMO detection based on mathematical concepts like lattice-reduction~(LR)~\cite{latRedux} or Monte-Carlo-Markov-Chain methods like Gibbs Sampling~\cite{gibbs}, but for wireless systems with large numbers of users and antennas, these conventional computing-based methods either suffer from drastic performance degradation or require an exponential increase in complexity to maintain BER performance. 

This prompts us to consider alternative computing approaches, specifically, the recent advancements in physics-inspired Ising machine-based computing methods like  quantum annealing~\cite{minsung}, Coherent Ising Machines~\cite{dopo,oeo}, digital-circuit Ising solvers~\cite{yamaoka201520k}, bifurcation machines~\cite{tatsumura2021scaling}, oscillation based Ising machines~\cite{oim}, and memristor/spintronic Ising machines~\cite{sutton2017intrinsic}. Recent advancements have shown promising results for solving tough problems in wireless communication. In particular, the QuAMax MIMO detector~\cite{minsung} was the first to use Quantum annealing as a physics-based Ising machine for MIMO detection. A classical-quantum hybrid approach~\cite{pSuccessQA} for MIMO detection can be used to further improve QA based MIMO detection. ParaMax~\cite{minsungParallelTemp} explores the use of parallel tempering for MIMO detection. Another exciting approach is the use of Oscillator-Ising machine, which is a purely classical system, for MU-MIMO detection in massive MIMO scenarios~\cite{oimMuMIMO}. QAVP~\cite{qavp} addresses downlink transmission in MU-MIMO systems. 

The paper introducing RI-MIMO~\cite{ri-mimo} was the first work exploring the use of Coherent Ising Machines for ML-MIMO detection, presenting a novel regularised Ising formulation for ML-MIMO in large and massive MIMO systems. 

\section{System Model}
\label{sec:sysModel}
Consider an uplink MIMO Scenario with $N_r$ antennas at the base station and $N_t$ users with one transmit antenna each. This is equivalent to a $N_t \times N_r$ MIMO system, and is described by an complex valued $N_r \times N_t$ channel matrix $\mathbf{\tilde{H}}$. The symbol transmitted by each user is drawn from a fixed set $\Phi$ representing the M-QAM constellation. Let the transmit vector be $\mathbf{\tilde{x}}$ ($\mathbf{\tilde{x}} \in \Phi^{N_t}$), then the received vector $\mathbf{\tilde{y}}$ is given by
\begin{equation}
    \mathbf{
    \tilde{y}} = \mathbf{\tilde{H}}\mathbf{\tilde{x}} + \mathbf{n},
\end{equation}
where $\mathbf{n}$ is the channel noise. If the channel noise is assumed to be white Gaussian noise then the maximum likelihood estimate for the transmitted vector $\mathbf{\tilde{x}}$ is given by
\begin{equation}
    \mathbf{\tilde{x}} = \arg \min_{\mathbf{\tilde{u}} \in \Phi^{N_t}} ||\mathbf{\tilde{y}}-\mathbf{\tilde{H}}\mathbf{\tilde{u}}||^2.
    \label{eq:ML-MIMO}
\end{equation}

Note that $\mathbf{\tilde{x}}$, $\mathbf{\textbf{y}}$ and $\mathbf{\textbf{H}}$ are all complex valued. Using the transform described in~\cite{ri-mimo},
\begin{equation}
\mathbf{H} = 
  \left[ {\begin{array}{cc}
   \Re(\mathbf{\tilde{H}}) & -\Im(\mathbf{\tilde{H}}) \\
   \Im(\mathbf{\tilde{H}}) & \Re(\mathbf{\tilde{H}}) \\
  \end{array} } \right],\text{~}
\end{equation}
\begin{equation}
\mathbf{y} =
  \left[ {\begin{array}{c}
   \Re(\mathbf{\tilde{y}}) \\
   \Im(\mathbf{\tilde{y}}) \\
  \end{array} } \right],\text{~}\mathbf{x}=
  \left[ {\begin{array}{c}
   \Re(\mathbf{\tilde{x}}) \\
   \Im(\mathbf{\tilde{x}}) \\
  \end{array} } \right],
  \label{eq:realTransVec}
\end{equation}
we get an equivalent real valued system,
\begin{equation}
    \mathbf{x} = \arg \min_{\mathbf{u} \in [\Re(\Phi)^{N_t},\Im(\Phi)^{N_t}]} ||\mathbf{y}-\mathbf{H}\mathbf{u}||^2.
    \label{eq:realML-MIMO}
\end{equation}
Our objective is to solve~(\ref{eq:realML-MIMO}), using a Coherent Ising Machine, and get the solution to~(\ref{eq:ML-MIMO}) by inverting the transform~(\ref{eq:realTransVec}). 

\section{Primer: Coherent Ising Machines}
\label{sec:cim}
An Ising minimization problem~\cite{isingNpHard1} is a quadratic unconstrained optimization problem given by
\begin{equation}
    \arg \min_{\forall i,s_i\in \{-1,1\}}-\sum_{i \neq j}J_{ij}s_{i}s_{j}=\min_{\mathbf{s}\in \{-1,1\}^N}  - \mathbf{s}^T\mathbf{J}\mathbf{s},
    \label{eq:Ising}
\end{equation}
where each \textit{spin} variable $s_{i} \in \{-1,1\}$, or in vector form (RHS)  $\mathbf{s} = \{s_1,s_2,...s_N\} \in \{-1,1\}^N$, 
where the diagonal entries of the matrix $\mathbf{J}$ are zeros. The Ising problem is known to be NP-Hard. 

In this paper, we focus on Coherent Ising Machines~(CIMs), originally proposed~(\cite{marandi2014network,mcmahon2016fully,inagaki2016coherent}) using an artificial optical spin network to find the ground state of the Ising problem. 
While CIMs principle of operation is centered around physics effects that are natively present in analog photonic hardware, using a network of degenerate optical parametric oscillators, in absence of quantum effects (which are very little significant in current hardware implementation), these machines can be emulated rather efficiently by a classical dynamical system description. Hence in this work we take the approach of considering the CIM a physics-based stochastic optimization algorithm that acts as a black box outputting a candidate solution of an Ising problem.

Numerically, the dynamics of the CIM can be modelled with real valued variables $x_i$ such that the corresponding \textit{spin} $s_i = sgn(x_i)$. Given the Ising optimization problem
\begin{equation}
    \arg \min_{s_1,s_2...s_N \in \{-1,1\}} -\sum_{j \neq i}J_{ij}s_is_j 
    \label{eq:IsingModel2}
\end{equation}

The time evolution of such a system can be modelled using the ODE~\cite{ampCorrectionCIM}
\begin{equation}
    \forall i\text{,      }\dfrac{dx_i}{dt} = (1-p)x_i - x_i^3 + \epsilon(t) \sum_{j\neq i} J_{i,j}x_j 
\end{equation}

Note that the CIM model used by RI-MIMO~\cite{ri-mimo} can be approximated by the above equation as well. An enhanced model with an amplitude heterogeneity correction~\cite{ampCorrectionCIM} can be used to destabilize the local minima and avoid limit cycles. The amplitude heterogeneity correction-based model introduces new ``error" variables $e_i$:
\begin{equation}
     \forall i \text{      ,}\dfrac{dx_i}{dt} = (1-p)x_i - x_i^3 + \epsilon(t) e_i \sum_{j\neq i} J_{i,j}x_j 
     \label{eq:cim1}
\end{equation}
\begin{equation}
     \forall i \text{,      } \dfrac{de_i}{dt} = -\beta(x_i^2 - a)e_i,\text{    }e_i > 0,
     \label{eq:cim2}
\end{equation}
where $a,p$, and $\beta$ are the free parameters of the model, and $\epsilon(t)$  is chosen to be a linear function of t, starting at zero and undergoing a slow ramp, $\epsilon=\gamma t$.  
\section{Design}
\label{sec:design}
The key idea of our proposed Ising MIMO mapping, which we call "Delta-Ising" (DI-MIMO), is to assume a guess for $\mathbf{x}$ in~(\ref{eq:realML-MIMO}) and optimize to find the best correction to the guess, rather than optimizing to find $\mathbf{x}$ itself (as existing state-of-the-art methods using Quantum Annealing or CIM do). 
 
We use the real valued equivalent of the MMSE estimate~($\mathbf{x_m}$) as the guess. This step has extremely low computation complexity and can be calculated from the MMSE solution using~(\ref{eq:realTransVec}). Given the solution variable $\mathbf{u}$ in the real-valued equivalent of the ML-MIMO problem~(\ref{eq:realML-MIMO}), we define
\begin{equation}
    \mathbf{d} =  \mathbf{u} - \mathbf{x}_{m},
    \label{eq:d_define}
\end{equation}
where $\mathbf{d}$ can be interpreted as the correction to the MMSE solution. The goal then becomes the formulation of an Ising problem for computing the optimal $\mathbf{d}$, such that $\mathbf{x}_{m} + \mathbf{d}$ will solve~(\ref{eq:realML-MIMO}). We use the MMSE solution as the guess, but another polynomial-time estimate of the MIMO problem, like FSD~\cite{fcsd} or SIC~\cite{sicMimo1} solution, could be used as well.

\begin{figure}[h!]
    \centering
    \includegraphics[width=\linewidth]{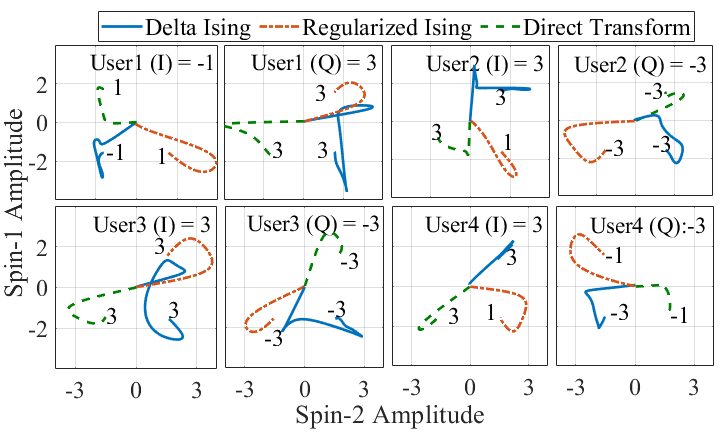}
    \caption{\textbf{Trajectories of Spin Amplitudes:} I-Q values of each user in an instance of 4$\times$4 MIMO, 16-QAM, and 10 dB SNR. We see that DI-MIMO can lead to different trajectories and final states than RI-MIMO and the direct Ising mapping. Note that the starting point of all trajectories is chosen, close to the origin, independently according to the distribution $\mathcal{N}(0,0.001)$.}
    \label{fig:IsingTrajectory10dB}
\end{figure}

For a QAM constellation, each element of $\mathbf{d}$ corresponds to the difference between the real or imaginary parts of two QAM symbols, and hence is an even valued integer. The choice of the search space for $\mathbf{d}$ is a design parameter; we can choose each element of $\mathbf{d}$ to belong to the set $\mathcal{D} = \{-2,0,2\}$, $\{-4,-2,0,2,4\}$, or an even larger search space. The choice of the search space for $\mathbf{d}$ reflects the region around the MMSE estimate that is searched for the optimal solution, and determines the Delta Ising formulation. Let $\Tilde{\mathbf{y}} = \mathbf{y}-\mathbf{H}\mathbf{x}_{m}$ be the residual received vector; then our new proposed formulation is obtained by substituting~(\ref{eq:d_define}) in~(\ref{eq:realML-MIMO})to change the optimization variable to $\mathbf{d}$, and changing the search space to $\mathcal{D}^{2\cdot N_t}$:
\begin{equation}
    \mathbf{\hat{d}} = \arg \min_{\mathbf{d} \in \mathcal{D}^{2\cdot N_t}} ||\Tilde{\mathbf{y}}-\mathbf{H}\mathbf{d}||^2.
    \label{eq:diIsing}
\end{equation}
The next step is to express $\mathbf{d}$ using \textit{spin} variables which can only take values $-1$ or $1$. Note that any scalar $c \in \{-2,0,2\}$ can be expressed using two \textit{spin} variables $s_1,s_2 \in \{-1,1\}$ as $c = s_1 + s_2$. Similarly, if $c \in \{-4,-2,0,2,4\}$ then it can be expressed using three \textit{spin} variables as $c = 2s_1+s_2+s_3$. Applying this reasoning for the vector $\mathbf{d}$, we can express $\mathbf{d}$ as
\begin{equation}
    \mathbf{d} = \mathbf{T}\mathbf{s},
    \label{eq:T_transform}
\end{equation}
 where $\mathbf{T}$ denotes a transform matrix that depends on the choice of $\mathcal{D}$, and $\mathbf{s}$ is a spin vector (each element is $\pm1$). 
 \begin{equation}
\mathbf{T} = 
  \left\{ {\begin{array}{cl}
    \text{ }[\mathbf{I}_{2\cdot N_t} \text{  }\mathbf{I}_{2\cdot N_t}],&  \mathcal{D} = \{-2,0,2\}\\
      \text{ }[2\mathbf{I}_{2\cdot N_t} \text{ }\mathbf{I}_{2\cdot N_t} \text{ }\mathbf{I}_{2\cdot N_t}],& \mathcal{D} = \{-4,-2,0,2,4\}\\
  \end{array} }\right.\text{.}
  \label{eq:T_def}
\end{equation}
If $\mathcal{D}$ is larger than $\{-4,-2,0,2,4\}$ then the corresponding $\mathbf{T}$ can be computed using similar reasoning as above. We get the equivalent Ising formulation of~(\ref{eq:realML-MIMO}) by substituting~(\ref{eq:T_transform}) in~(\ref{eq:diIsing}):
 \begin{equation}
     \hat{\mathbf{s}} = \arg \min_{\mathbf{s} \in \{-1,1\}^{2\cdot N_t}} -\mathbf{h}^T\mathbf{s} - \mathbf{s}^T\mathbf{J}\mathbf{s},
     \label{eq:IsingMIMO}
 \end{equation}
 \begin{equation}
     \mathbf{J} = -zeroDiag(\mathbf{T'H'HT})\text{ ,  } \mathbf{h} = 2\Tilde{\mathbf{y}'}\mathbf{HT},
     \label{eq:IsingCoeff}
 \end{equation}
where $zeroDiag(.)$ sets the diagonal elements to zero. We solve the obtained Ising problem, using an Ising machine, to get $\mathbf{\hat{s}}$; and then compute $\hat{\mathbf{d}} = \mathbf{T}\mathbf{\hat{s}}$ and $\mathbf{x} = \mathbf{x_m} + \mathbf{\hat{d}}$. Since some elements of $\mathbf{x}$ can be outside the QAM constellation, we round $\mathbf{x}$ to the nearest QAM symbol. Finally, we can get the maximum likelihood MIMO solution, $\mathbf{\Tilde{x}}$ in~(\ref{eq:ML-MIMO}), from $\mathbf{x}$ by inverting the transform described in~(\ref{eq:realTransVec}). 
\\\\
The Ising problem obtained in~(\ref{eq:IsingMIMO}) has both linear and quadratic terms; however, as we saw in Section~\ref{sec:cim}, to maintain compatibility with optical hardware implementations, the CIM requires an Ising problem with only quadratic terms. We use an auxiliary spin variable~\cite{ri-mimo} to convert all linear terms in~(\ref{eq:IsingMIMO}) to quadratic and define an auxiliary Ising problem:
 \begin{equation}
     (\bar{\mathbf{s}},\bar{s_a}) = \arg \min_{\mathbf{s} \in \{-1,1\}^{2N_t},s_a\in\{-1,1\}} -(\mathbf{h}^T\mathbf{s})s_a - \mathbf{s}^T\mathbf{J}\mathbf{s}.
 \end{equation}
and the solution of~(\ref{eq:IsingMIMO}) can be obtained from the solution of the auxiliary Ising problem~\cite{ri-mimo}  using the equation $\hat{\mathbf{s}} = \bar{\mathbf{s}}\times \bar{s_a}$. This allows us to solve~(\ref{eq:IsingMIMO}) on the CIM. Our new formulation changes the time evolution trajectories of the CIM and we will demonstrate this later in Fig.~\ref{fig:IsingTrajectory10dB}.
\\
We solve each problem on the amplitude heterogeneity correction-based CIM (described by~(\ref{eq:cim1}) and~(\ref{eq:cim2})). Borrowing the terminology from the Quantum Annealing literature, we refer to each run on the CIM as  an ``anneal.'' A standard practice is to run multiple anneals, for the same problem, to enhance the probability of finding the optimal solution. Hence, we run $N_a$ anneals for each problem instance, \textit{i.e.}, we solve each problem $N_a$ times on the CIM and generate $N_a$ candidate solutions. Then we compare the obtained candidate solutions and the MMSE solution and select the best.

% While it might seem that the Delta Ising formulation is a linear translation of the optimization problem, along with a restriction on the search space; it is important to note that the Ising coefficients generated by the Delta-Ising formulation in~(\ref{eq:IsingCoeff}) are very different. Hence, due to the non-linear dynamics of the CIM (described by~(\ref{eq:cim1}) and~(\ref{eq:cim2})), the time evolution trajectories of the CIM are significantly different and we empirically observe an overall increase in quality of the solutions found by the CIM, which is reflected through a superior BER performance. We later will demonstrate the variation in time evolution trajectories in Fig.~\ref{fig:IsingTrajectory10dB}. 

% \begin{figure*}[h!]
%     \centering
%     \includegraphics[width=0.9\linewidth]{figures/IsingTrajectories_noNoise.png}
%     \caption{Trajectories of Spin Amplitudes for I-Q values of each users in an instance of 4$\times$4 MIMO, 16 QAM and no noise.}
%     \label{fig:IsingTrajectoryNoNoise}
% \end{figure*}

For illustrative purposes, we compare the direct ML-MIMO-to-Ising transformation initially introduced by QuAMax~\cite{minsung} (which involves representing the I/Q components of the QAM constellation using \textit{spin} variables), RI-MIMO~\cite{ri-mimo} (which modifies the direct transformation by adding a suitable regularisation term) and the DI-MIMO. We present one of the time evolution of state variables of the CIM described by (\ref{eq:cim1}) and (\ref{eq:cim2}) and show that DI-MIMO can lead to very different trajectories and final states. We consider decoding instances from a $4\times4$ MIMO system with 16-QAM modulation, and $\mathcal{D} = \{-2,0,2\}$ for DI-MIMO. All three formulations use two spins to represent the I/Q components of the QAM constellation. Observe the time evolution of the CIM state variables corresponding to the two spin variables (for each I/Q component) at 10~dB SNR~(Fig.~\ref{fig:IsingTrajectory10dB}). The Ising coefficients of the  DI formulation in~(\ref{eq:IsingCoeff}) are very different compared to RI-MIMO and direct formulations, further contributing to the distinct dynamics of (\ref{eq:cim1}).

\section{Evaluation}
\label{sec:eval}
In this section, we measure the BER and spectral efficiency achieved by DI-MIMO and compare to RI-MIMO and MMSE. 

\subsection{Implementation of the CIM Simulator}
In this work, we implement an Ising solver by simulating the time evolution of the amplitude heterogeneity correction-based bi-stable system described by~(\ref{eq:cim1}) and~(\ref{eq:cim2}). Our simulator performs numerical integration with $dt = 0.01$ for 256 integration steps. The initial values $x_i$ are i.i.d $\mathcal{N}(0,0.001)$ and $e_i$ are initialized using a folded $\mathcal{N}(0,0.001)$ distribution. Inspired by the results in \cite{mcmahon2016fully}, we set $p = 0.98$, $\beta = 1$, $a = 2$ and $\gamma = 1000/(256\cdot0.01)$. These parameters were empirically selected, based on trial-and-error experiments, such that the system can achieve steady state and provide good performance. Note that performance can be further improved by optimally selecting these parameters, as well as adding post-processing improvements, and we plan to address this in our future work. For instance, we use only the final equilibrium state (the value of $x_i$ at the end of 256 integration steps) of the amplitude correction based CIM to calculate the candidate solution, instead of comparing against all intermediate CIM states during the computation.
\begin{figure*}[ht!]
    \centering
    \includegraphics[width=\linewidth]{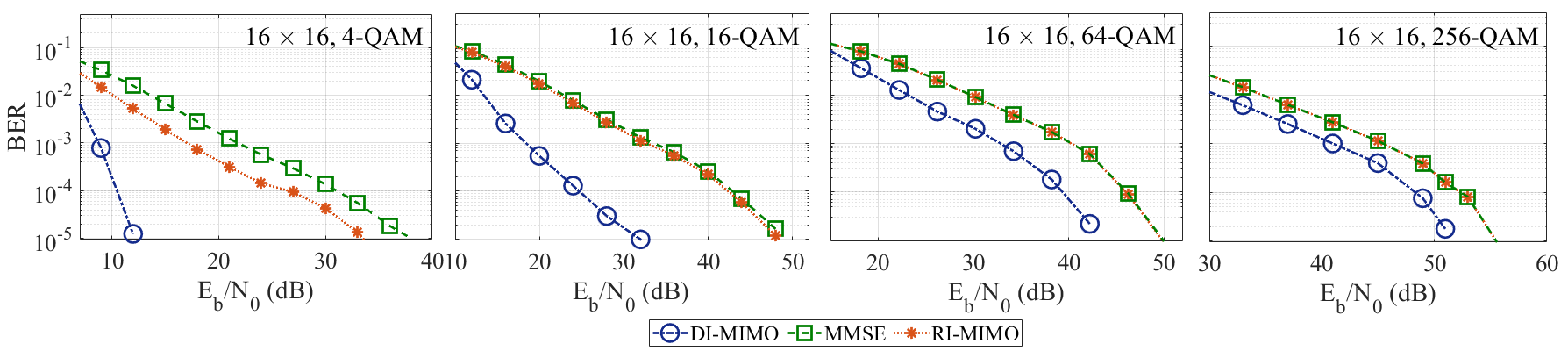}
    \caption{\textbf{BER (Large MIMO configuration)} of the proposed DI-MIMO for $16\times16$ Large MIMO system with 4-, 16-, 64-, and 256-QAM modulation. While RI-MIMO struggles to improve MMSE performance for 16-QAM or higher, DI-MIMO improves BER even at high modulation orders.}
    \label{fig:ber_plot_large}
\end{figure*}
\begin{figure*}[ht!]
    \centering
    \includegraphics[width=\linewidth]{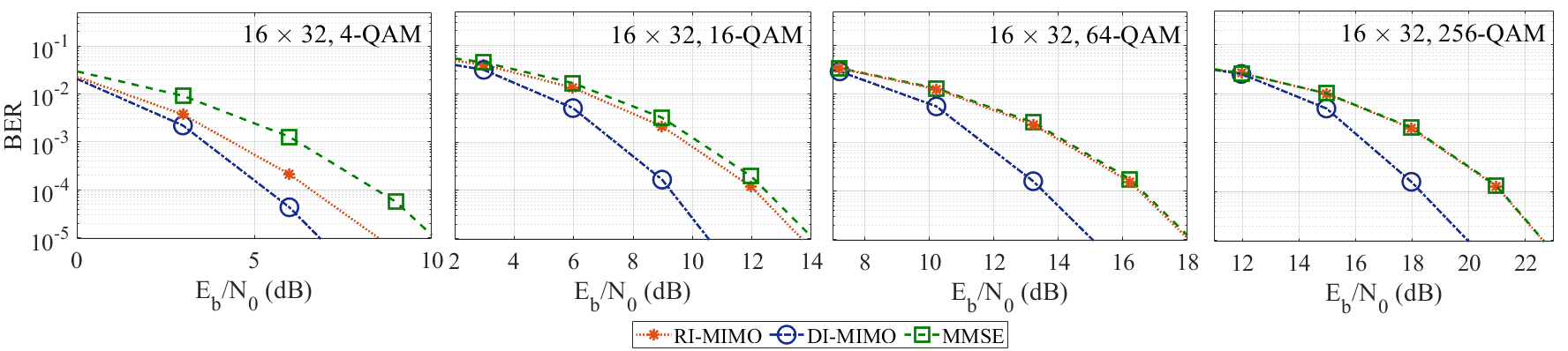}
    \caption{\textbf{BER (Massive MIMO configuration)} of the proposed DI-MIMO for $16\times32$ Massive MIMO systems with 4-, 16-, 64-, and 256-QAM modulation. While RI-MIMO struggles to improve MMSE performance for 16-QAM or higher, DI-MIMO improves BER even at high modulation orders.}
    \label{fig:ber_plot_massive}
\end{figure*}
\subsection{Evaluation Setup}
Our evaluation setup simulates an uplink $N_u\times N_r$ MIMO system which has $N_u$ users (with one transmit antenna each) and $N_r$ receive antennas at the base station ($N_r > N_u$). We assume a slow fading channel and channel instances are assumed to follow the Rayleigh fading model. We simulate the transmission of a total of $2\times10^{6}$ bits, equally divided between the users, and the BER is computed as the mean BER of all users. We compare our methods against RI-MIMO~\cite{ri-mimo}, which also targets MIMO detection on CIMs, and the MMSE detector, in both large and massive MIMO scenarios. 

\subsection{BER Performance}
We first simulate the BER performance of our methods for a $16\times16$  large MIMO system with 4-, 16-, 64-, and 256-QAM modulations. We select the search space around the MMSE solution to be $\mathcal{D} = \{-2,0,2\}$, corresponding to search across the immediately neighboring 8 symbols for each user, and execute $N_a = 64$ anneals per instance ( inspired by the setup of \cite{ri-mimo}). We see from Fig.~\ref{fig:ber_plot_large} that DI-MIMO provides significant performance gains over MMSE and RI-MIMO~(which struggles to improve MMSE performance for 16-QAM or higher). We see that for 4- and 16-QAM modulation, DI-MIMO provides about 20~dB gain over RI-MIMO/MMSE. While the performance of RI-MIMO is nearly the same as MMSE for 16-QAM or higher modulation, DI-MIMO provides significant BER improvements even at 256-QAM. The inability to improve the performance~(with a reasonable number of anneals) of MMSE (for 16-QAM and higher modulation) is the key limitation of RI-MIMO~\cite{ri-mimo}. Fig.~\ref{fig:ber_plot_large} demonstrates that DI-MIMO is capable of overcoming this limitation and extending the benefits of CIM-based MIMO detection to very large modulations like 256-QAM.

Next, let us look at BER performance for massive MIMO systems, which have a much higher number of receive antennas at the base station than users/transmit antennas. Massive MIMO systems have highly well-conditioned channels, and even linear methods like MMSE can achieve low BERs~\cite{massiveMimoNearOpt}.
We simulate the BER performance for a $16\times32$ massive MIMO system:
 we see~(Fig.~\ref{fig:ber_plot_massive}) similar performance trends as large MIMO. DI-MIMO provides significant performance gains over RI-MIMO/MMSE (for even 256-QAM), whereas RI-MIMO struggles to improve MMSE beyond 4-QAM. 
\begin{figure}[h!]
    \centering
    \includegraphics[width=\linewidth]{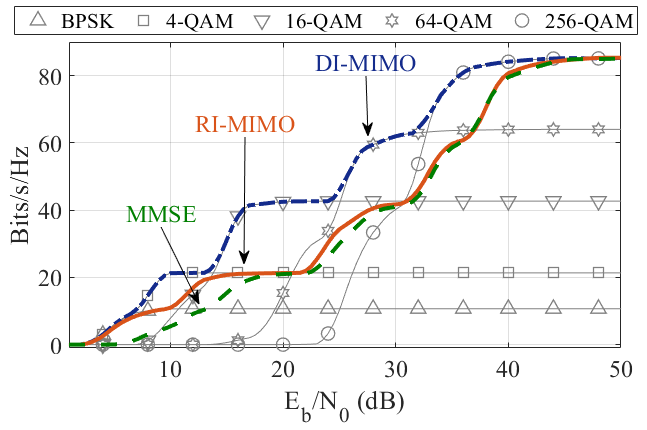}
    \caption{\textbf{Large MIMO spectral efficiency:} $16\times16$ MIMO system with Adaptive Modulation and Coding: demonstrating that the proposed DI-MIMO can significantly improve MMSE and RI-MIMO performance. Note that RI-MIMO improves MMSE only for BPSK and 4-QAM. The faint grey curves correspond to the throughput of DI-MIMO with fixed modulation and adaptive selection of code rate. The envelope of these grey curves (blue curve) is the spectral efficiency of DI-MIMO with Adaptive Modulation and Coding.} 
    \label{fig:tputLarge}
% \end{figure}
% \begin{figure}[t!]
    \centering
    \includegraphics[width=\linewidth]{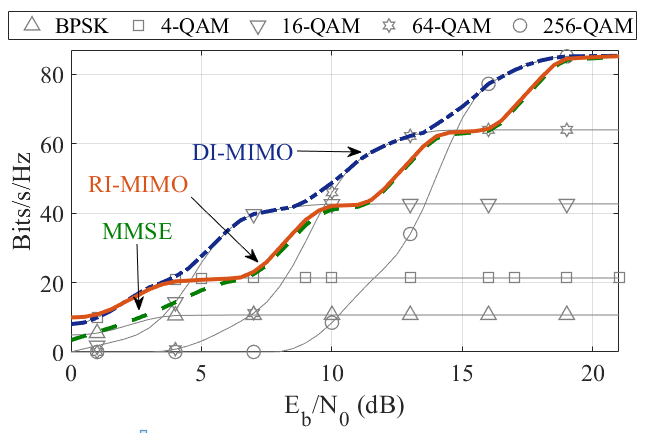}
    \caption{\textbf{Massive MIMO spectral efficiency:} $16\times32$ MIMO system with Adaptive Modulation and Coding: demonstrating that the proposed DI-MIMO can significantly improve MMSE and RI-MIMO performance. Note that RI-MIMO improves MMSE only for BPSK and 4-QAM. The faint grey curves correspond to the throughput of DI-MIMO with fixed modulation and adaptive selection of code rate. The envelope of these grey curves (blue curve) is the spectral efficiency of DI-MIMO with Adaptive Modulation and Coding.}
    \label{fig:tputMassive}
\end{figure}
\subsection{Spectral Efficiency}
We now turn to spectral efficiency: in a real-life wireless system, the modulation and coding rate of transmission are selected based on the SNR to maximize spectral efficiency by an Adaptive Modulation and Coding~(AMC) function. To avoid the complexities of implementing AMC, we assume that we have an oracle AMC function that selects the best modulation and code rate at any given SNR. We allow our system to use convolutional coding with code rates of $\frac{1}{3}$, $\frac{1}{2}$, and $\frac{2}{3}$; and use BPSK, 4-QAM, 16-QAM, 64-QAM, and 256-QAM modulations. We report the simulated spectral efficiency of a $16\times16$ MIMO system in Fig~\ref{fig:tputLarge}. Note that DI-MIMO (with only $N_a= 64$ anneals per instance) can provide significant throughput gains over MMSE, achieving $2\times$ throughput for SNR $\leq25$ dB, and up to $1.5\times$ throughput for SNR $\geq 30$ dB. In contrast, note that RI-MIMO only improves the MMSE performance in low SNR regimes when the AMC function is using BPSK or 4-QAM modulation, and its performance is similar to MMSE when 16-QAM or higher is used. 

In the massive MIMO regime, we simulate the spectral efficiency of a $16\times32$ MIMO system in Fig.~\ref{fig:tputMassive}. We see similar performance trends as the $16\times16$ MIMO system. DI-MIMO can provide significant throughput gains over MMSE/RI-MIMO, achieving $2\times$ throughput in the mid-SNR regime ($\approx 7.5$ dB) and $1.5\times$ throughput in the high-SNR regime~($\approx 12$ dB).

\section{Conclusion and Future Work}
\label{sec:conclusion}
In this paper, we have presented a novel Ising formulation for Maximum Likelihood MIMO detection and explored the use of an enhanced model of the Coherent Ising Machine. Our proposed DI-MIMO algorithm converts the ML-MIMO problem into a perturbative correction to a fast candidate decoded MMSE solution and solves it on the amplitude heterogeneity correction-based CIM. We show that, unlike the existing Ising-machine-based MIMO detectors, which struggle with 16-QAM or higher modulations, DI-MIMO can provide significant performance gains with high modulation orders.

Unlike the existing state-of-the-art for Ising Machine-based MIMO detection, the transform used by the DI formulation depends only on the choice of search space and is independent of modulation. Hence, we plan to explore the design of highly optimized fixed FPGA/ASIC-based implementations for practical systems where the modulation is adaptive and is varied to maximize spectral efficiency. 


\section*{Acknowledgments}

D.V. acknowledges support from NSF CNS-1824470 and CCF-1918549.
A.K.S. and K.J. acknowledge support from NSF CNS-1824357.

\bibliographystyle{IEEEtran}
\bibliography{IEEEabrv,ref}

\begin{thebibliography}{10}
\providecommand{\url}[1]{#1}
\csname url@samestyle\endcsname
\providecommand{\newblock}{\relax}
\providecommand{\bibinfo}[2]{#2}
\providecommand{\BIBentrySTDinterwordspacing}{\spaceskip=0pt\relax}
\providecommand{\BIBentryALTinterwordstretchfactor}{4}
\providecommand{\BIBentryALTinterwordspacing}{\spaceskip=\fontdimen2\font plus
\BIBentryALTinterwordstretchfactor\fontdimen3\font minus
  \fontdimen4\font\relax}
\providecommand{\BIBforeignlanguage}[2]{{%
\expandafter\ifx\csname l@#1\endcsname\relax
\typeout{** WARNING: IEEEtran.bst: No hyphenation pattern has been}%
\typeout{** loaded for the language `#1'. Using the pattern for}%
\typeout{** the default language instead.}%
\else
\language=\csname l@#1\endcsname
\fi
#2}}
\providecommand{\BIBdecl}{\relax}
\BIBdecl

\bibitem{mimoSurveyLarge}
S.~Yang and L.~Hanzo, ``{Fifty Years of MIMO Detection: The Road to Large-Scale
  MIMOs},'' \emph{{IEEE Communications Surveys Tutorials}}, 2015.

\bibitem{sphereIeee}
C.~{Hung} and T.~{Sang}, ``{A Sphere Decoding Algorithm for MIMO Channels},''
  in \emph{2006 IEEE International Symposium on Signal Processing and
  Information Technology}, 2006, pp. 502--506.

\bibitem{sphereComp}
B.~{Hassibi} and H.~{Vikalo}, ``{On the sphere-decoding algorithm I. Expected
  complexity},'' \emph{IEEE Transactions on Signal Processing}, 2005.

\bibitem{mohseni2022ising}
N.~Mohseni, P.~L. McMahon, and T.~Byrnes, ``Ising machines as hardware solvers
  of combinatorial optimization problems,'' \emph{arXiv preprint
  arXiv:2204.00276}, 2022.

\bibitem{ri-mimo}
\BIBentryALTinterwordspacing
A.~K. Singh, K.~Jamieson \emph{et~al.}, ``Ising machines' dynamics and
  regularization for near-optimal large and massive mimo detection.''\hskip 1em
  plus 0.5em minus 0.4em\relax arXiv, 2021. [Online]. Available:
  \url{https://arxiv.org/abs/2105.10535}
\BIBentrySTDinterwordspacing

\bibitem{minsung}
M.~Kim \emph{et~al.}, ``{Leveraging Quantum Annealing for Large MIMO Processing
  in Centralized Radio Access Networks},'' \emph{{The 31st ACM Special Interest
  Group on Data Communication (SIGCOMM)}}, 2019.

\bibitem{minsungParallelTemp}
M.~Kim, S.~Mandra, D.~Venturelli, and K.~Jamieson, ``{Physics-Inspired
  Heuristics for Soft MIMO Detection in 5G New Radio and Beyond},'' in
  \emph{Proceedings of the 27th Annual International Conference on Mobile
  Computing and Networking (MobiCom)}, 2021.

\bibitem{pSuccessQA}
\emph{{Towards Hybrid Classical-Quantum Computation Structures in
  Wirelessly-Networked Systems}}, 2020.

\bibitem{leleu2019destabilization}
T.~Leleu, Y.~Yamamoto, P.~L. McMahon, and K.~Aihara, ``{Destabilization of
  local minima in analog spin systems by correction of amplitude
  heterogeneity},'' \emph{Physical review letters}, vol. 122, no.~4, p. 040607,
  2019.

\bibitem{sicMimo1}
C.~{Navarro Manchon}, L.~{Deneire}, P.~{Mogensen}, and T.~B. {Sorensen}, ``{On
  the Design of a MIMO-SIC Receiver for LTE Downlink},'' in \emph{2008 IEEE
  68th Vehicular Technology Conference}, 2008, pp. 1--5.

\bibitem{fcsd}
L.~G. {Barbero} and J.~S. {Thompson}, ``{Fixing the Complexity of the Sphere
  Decoder for MIMO Detection},'' \emph{IEEE Transactions on Wireless
  Communications}, vol.~7, no.~6, pp. 2131--2142, 2008.

\bibitem{latRedux}
B.~{Gestner}, W.~{Zhang} \emph{et~al.}, ``{Lattice Reduction for MIMO
  Detection: From Theoretical Analysis to Hardware Realization},'' \emph{IEEE
  Transactions on Circuits and Systems I: Regular Papers}, 2011.

\bibitem{gibbs}
A.~M. Mussi and T.~Abrão, ``\BIBforeignlanguage{English}{Multiple restarts
  mixed gibbs sampling detector for large-scale antenna systems},''
  \emph{\BIBforeignlanguage{English}{IET Signal Processing}}, 2019.

\bibitem{dopo}
Y.~Haribara, S.~Utsunomiya, and Y.~Yamamoto, ``{Computational Principle and
  Performance Evaluation of Coherent Ising Machine Based on Degenerate Optical
  Parametric Oscillator Network},'' \emph{Entropy}, 2016.

\bibitem{oeo}
F.~Böhm, G.~Verschaffelt, and G.~Van~der Sande, ``{A poor man’s coherent
  Ising machine based on opto-electronic feedback systems for solving
  optimization problems},'' \emph{Nature Communications}, vol.~10, 2019.

\bibitem{yamaoka201520k}
M.~Yamaoka, C.~Yoshimura \emph{et~al.}, ``{A 20k-spin Ising chip to solve
  combinatorial optimization problems with CMOS annealing},'' \emph{IEEE
  Journal of Solid-State Circuits}, 2015.

\bibitem{tatsumura2021scaling}
K.~Tatsumura \emph{et~al.}, ``{Scaling out Ising machines using a multi-chip
  architecture for simulated bifurcation},'' \emph{Nature Electronics}, 2021.

\bibitem{oim}
T.~Wang and J.~Roychowdhury, ``{OIM: Oscillator-Based Ising Machines for
  Solving Combinatorial Optimisation Problems},'' in \emph{Unconventional
  Computation and Natural Computation}, I.~McQuillan and S.~Seki, Eds.\hskip
  1em plus 0.5em minus 0.4em\relax Cham: Springer International Publishing,
  2019, pp. 232--256.

\bibitem{sutton2017intrinsic}
B.~Sutton, K.~Y. Camsari, B.~Behin-Aein, and S.~Datta, ``{Intrinsic
  optimization using stochastic nanomagnets},'' \emph{Scientific reports},
  2017.

\bibitem{oimMuMIMO}
J.~Roychowdhury, J.~Wabnig, K.~P. Srinath \emph{et~al.}, ``{Performance of
  Oscillator Ising Machines on Realistic MU-MIMO Decoding Problems, 22
  September 2021, PREPRINT (Version 1)}.''

\bibitem{qavp}
S.~Kasi, A.~K. Singh \emph{et~al.}, ``{Quantum Annealing for Large MIMO
  Downlink Vector Perturbation Precoding},'' in \emph{Forthcoming, IEEE
  International Conference on Communications (ICC)}, 2021.

\bibitem{isingNpHard1}
A.~Lucas, ``{Ising formulations of many NP problems},'' \emph{Frontiers in
  Physics}, vol.~2, p.~5, 02 2014.

\bibitem{marandi2014network}
A.~Marandi, Z.~Wang, K.~Takata, R.~L. Byer, and Y.~Yamamoto, ``{Network of
  time-multiplexed optical parametric oscillators as a coherent Ising
  machine},'' \emph{Nature Photonics}, 2014.

\bibitem{mcmahon2016fully}
P.~L. McMahon, A.~Marandi \emph{et~al.}, ``{A fully programmable 100-spin
  coherent Ising machine with all-to-all connections},'' \emph{Science}, 2016.

\bibitem{inagaki2016coherent}
T.~Inagaki, Y.~Haribara \emph{et~al.}, ``{A coherent Ising machine for
  2000-node optimization problems},'' \emph{Science}, vol. 354, no. 6312, pp.
  603--606, 2016.

\bibitem{ampCorrectionCIM}
T.~Leleu, Y.~Yamamoto, P.~L. McMahon, and K.~Aihara, ``Destabilization of local
  minima in analog spin systems by correction of amplitude heterogeneity,''
  \emph{Phys. Rev. Lett.}, vol. 122, p. 040607, Feb 2019.

\bibitem{massiveMimoNearOpt}
M.~{Mandloi} and V.~{Bhatia}, ``{Low-Complexity Near-Optimal Iterative
  Sequential Detection for Uplink Massive MIMO Systems},'' \emph{IEEE
  Communications Letters}, vol.~21, no.~3, pp. 568--571, 2017.

\end{thebibliography}

\end{document}